\documentclass[a4paper,11pt]{revtex4}
\usepackage{amsmath}
\usepackage{amssymb}
\usepackage{amscd}
\usepackage{graphicx}
\usepackage{bbm}
\usepackage{hyperref}

\begin{document}
\title{Knotted Defects in Nematic Liquid Crystals}
\author{Thomas Machon}
\author{Gareth P. Alexander}
\affiliation{Department of Physics and Centre for Complexity Science, University of Warwick, CV4 7AL, UK}

\date{\today}

\begin{abstract}
We show that the number of distinct topological states associated to a given knotted defect, $L$, in a nematic liquid crystal is equal to the determinant of the link $L$. 
We give an interpretation of these states, demonstrate how they may be identified in experiments and describe the consequences for material behaviour and interactions between multiple knots. 
We show that stable knots can be created in a bulk cholesteric and illustrate the topology by classifying a simulated Hopf link. 
In addition we give a topological heuristic for the resolution of strand crossings in defect coarsening processes which allows us to distinguish topological classes of a given link and to make predictions about defect crossings in nematic liquid crystals.
\end{abstract}
\maketitle

Topological concepts have come to play an increasingly significant role in characterising and controlling material behaviour across all areas of condensed matter, encompassing vortices in fluids~\cite{woltjer58,moffatt69}, defects in ordered media~\cite{toulouse76,volovik76,kleman77}, the quantum Hall effect~\cite{vonKlitzing80,thouless82}, colloids in liquid crystals~\cite{terentjev95,poulin97,musevic06,senyuk13}, topological insulators~\cite{kane05,hasan10}, and boundary modes in isostatic lattices~\cite{sun12,kane13}. A recent development has been the experimental creation of knotted field configurations in laser light~\cite{dennis10}, liquid crystals~\cite{tkalec11,chen13,senyuk13,cavallaro13,martinez14} and fluid flows~\cite{kleckner13}. In liquid crystals knotted fields are often produced, or controlled, by colloidal particles and their associated topological defects~\cite{tkalec11,senyuk13,machon13}. By suitable manipulation with laser tweezers, the defect lines can be tied into arbitrary knots and links~\cite{tkalec11,kamien11}. Furthermore, modern fabrication techniques allow the colloids themselves to be made in the shape of a knot~\cite{martinez14}, including with hybrid surface anchoring conditions~\cite{cavallaro13} so that the colloid faithfully mimics a defect line. Thus more or less arbitrary knotted textures can be made and manipulated and it is important to understand and characterise their properties. 

Knots are intricate entities that display enormous diversity. Different knots, and links, distinguish themselves through a large number of topological invariants, including various polynomials, homology groups and homotopy groups~\cite{Rolfsen,Lickorish}. These must be encoded in any knotted field, although, at present, it is almost completely unknown which invariants manifest themselves in distinct material properties or behaviour. A most basic question is to determine how many topologically distinct knotted configurations are compatible with a given link, $L$, and characterise them. For instance, there are an infinite number of distinct point vortices in the XY model labelled by an integer winding number, and the distinct types of topological insulators and superconductors can be classified by a periodic table using Bott periodicity~\cite{kitaev09}. Here we show that the number of distinct knotted fields in a nematic liquid crystal (up to homotopy) is equal to the determinant of the link, {\it i.e.} the order of the first homology of the double branched cover of the link complement. This leads to a topological characterisation of the interaction between knotted structures in nematics based on the known properties of link determinants and an understanding of how defect lines reconfigure following strand crossings. 

Defect lines in nematics, whether genuine or imitated by colloids~\cite{cavallaro13}, are regions where the material order is undefined and are characterised by the property that the molecular alignment reverses orientation upon going around the line {\sl once}~\cite{alexander12}. This gives a vivid visual demonstration that the order in nematics corresponds to a unit line field, called the director, rather than a vector. Nonetheless, it is a widely adopted convention to treat the director as a vector field and impose the equivalence $n\sim-n$ `by hand'. A formal treatment of this process provides a natural framework for characterising knotted nematics. Since the alignment returns to the same orientation upon going around the disclination {\sl twice}, any knotted director field, $n$, can be lifted to a vector field, $\hat{n}$, on the cyclic double cover $\widehat{\Sigma}(L)$ of the link complement, a process analogous with the familiar branch cuts of complex analysis. It is the topology of this space that is explored by the nematic, which provides a physical realisation of the double branched cover of $S^3$ over the link. 

After passing to the vector field $\hat{n}$, the equivalence relation $n \sim -n$ is restated as a compatibility condition with respect to the deck transformation, $t$, of the covering space (that moves us from a point $x$ on one sheet of the cover to the equivalent point of the other)
\begin{equation}
\hat{n}(tx) = - \hat{n}(x). 
\end{equation}
Such vector fields are said to be equivariant. So the classification of knotted nematics with a given link as defect set can be equivalently posed as the classification of equivariant vector fields on the cyclic double cover of the link complement up to equivariant homotopy. As is common in experimental systems, we take the director to be uniform at large distances, and thus the conventional link complement $S^3 - L$ as our domain.

\begin{figure}
\begin{center}
\includegraphics{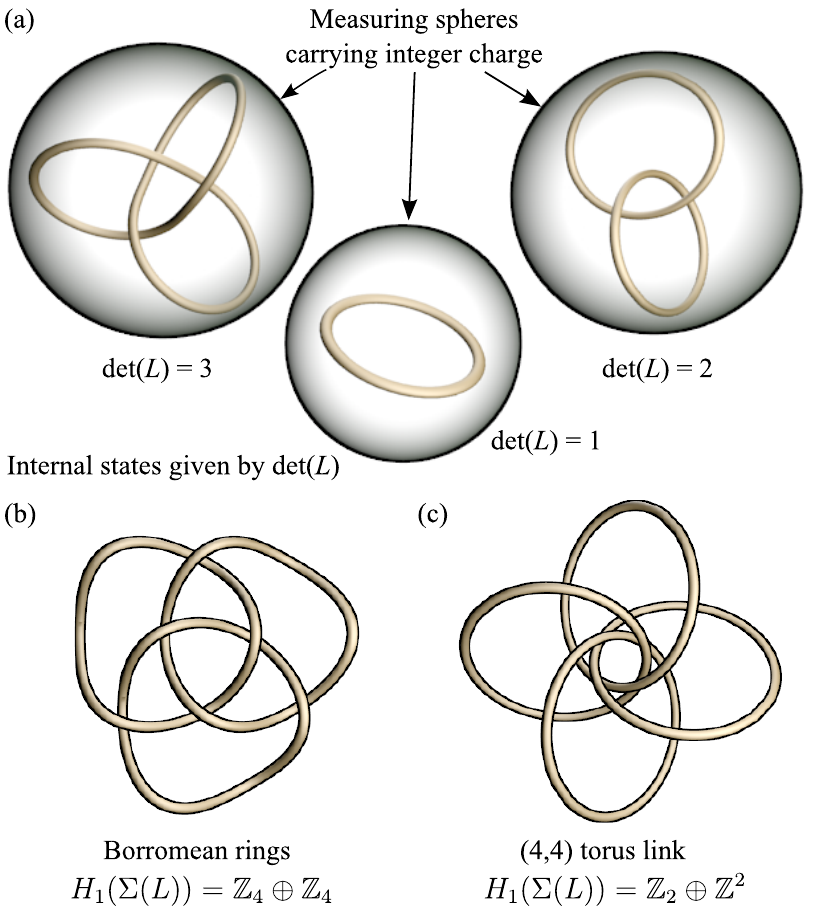}
\end{center}
\caption{(a) `Particle' based picture of knotted defects in liquid crystals. Each link (L-R: trefoil knot, unknot and Hopf link) has an internal degree of freedom given by an element of $H_1(\Sigma(L))$, the size of which is $\det (L)$. Each split component then carries a hedgehog charge, calculated using the degree of the texture on the measuring spheres~\cite{mermin79,alexander12}, which is constrained by charge conservation. (b) Borromean rings; for this link $H_1(\Sigma(L))=\mathbb{Z}_4 \oplus \mathbb{Z}_4$ giving a total of 16 distinct states. (c) (4,4) torus link; this link has $H_1(\Sigma(L))=\mathbb{Z}_2 \oplus \mathbb{Z}^2$, and thus supports an infinite number of states.}
\label{fig:1}
\end{figure}

Without imposing equivariance, unit vector fields on the cyclic double cover are classified by the induced map on second cohomology and hence by the group $H^2(\widehat{\Sigma}(L))$~\cite{whitney37}. By Poincar\'{e}-Lefschetz duality this is isomorphic to the relative homology group $H_1( \widehat{\Sigma}(L),\partial \widehat{\Sigma}(L))$, the elements of which are cycles that `entangle' the knot and `tethers' that connect various link components. Restricting to equivariant maps only allows cycles of the form $(e-te)$. Equivalence between a pair of cycles $(e-te)$ and $(e^\prime-t e^\prime)$ is established by considering equivariant homotopies that exchange cycles across any branching surface between the two sheets of $\widehat{\Sigma}(L)$. These relations reduce the group $H_1( \widehat{\Sigma}(L),\partial \widehat{\Sigma}(L))$ to $H_1(\Sigma(L))$, the first homology group of the double branched cover of $S^3$ over $L$. The order of this group (if finite) is known as the knot determinant, a well-known and effectively computable knot invariant, and counts the number of topologically distinct nematic textures associated to a given link, $L$~\cite{note1}. For example, the number of unknot textures is one, the number of trefoil knots is three and the number of Hopf links is two, shown in Fig.~\ref{fig:1}. There are also links that support an infinite number of states, the $(4,4)$ torus link (see Fig.~\ref{fig:1}(c)), for example, has $H_1(\Sigma(L))= \mathbb{Z}_2 \oplus \mathbb{Z}^2$ meaning that the state is described by three integers, one of which is defined only mod 2. 

We can use this result to understand the topology of multiple knotted and linked defects in a nematic. If a given link $L$ is split (meaning that it has multiple components that can each be surrounded by a measuring sphere in the space) into say $L_1, \dots, L_n$, shown in Fig.~\ref{fig:1}(a), then $H_1(\Sigma(L))$ splits as a direct sum~\cite{Rolfsen} 
\begin{equation}
H_1\bigl(\Sigma(L)\bigr)=\mathbb{Z}^{n-1}\oplus \left(\bigoplus_{i=1}^n H_1\bigl(\Sigma(L_i)\bigr) \right).
\label{eq:1}
\end{equation}
This equation encodes the topological interaction between a collection of knots and links. Indeed, one can think of each split component as a knotted `particle', the internal states of which are given by $H_1(\Sigma(L_i))$, \textit{i.e.} the determinant of the component on its own. This interpretation, reminiscent of Kelvin's `vortex atoms'~\cite{thomson67}, is supplemented by a topological interaction between the components. This is specified by an integer associated to each component, interpreted as the usual `hedgehog charge' that identifies point defects. This gives the factor of $\mathbb{Z}^{n-1}$ in~\eqref{eq:1}, there being only $n-1$ degrees of freedom due to the conserved total charge imposed by the uniform far-field boundary conditions. Each factor of $\mathbb{Z}$ can be calculated in the usual way by considering the director field on a measuring sphere and computing the degree of the map~\cite{mermin79,alexander12}.

\begin{figure}
\begin{center}
\includegraphics{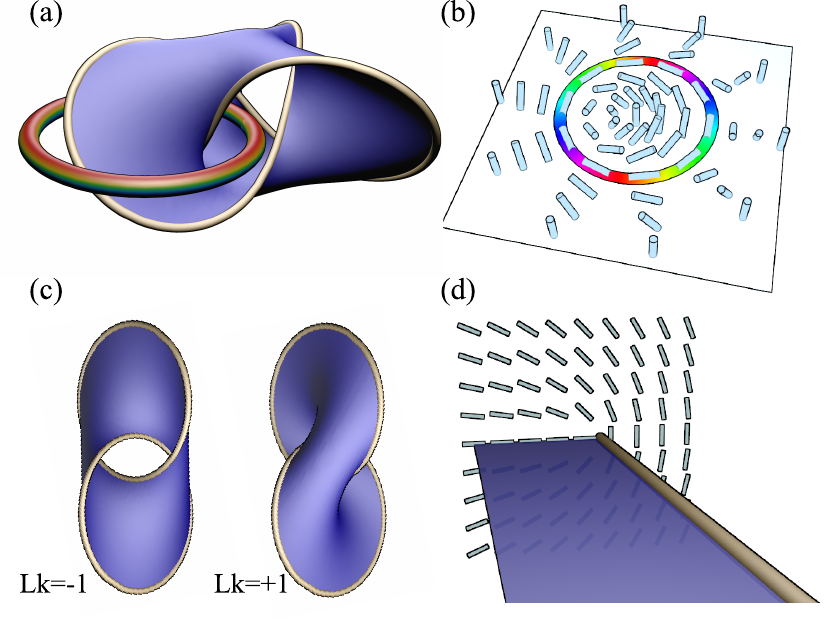}
\end{center}
\caption{Using the Pontryagin-Thom construction to understand knotted textures. (a) Cartoon PT surface (blue) for a trefoil knot, entangled with a Skyrmion tube. The tube cannot be removed without altering the topology of the defect line; the number of these tubes distinguishes the different topological classes. The colouring of the Skyrmion tube corresponds to the colouring of the strip in (b). (b) Cartoon showing the relation between a Skyrmion (tube) and the PT surface. The surface is drawn where the director is horizontal, and coloured by orientation as shown. (c) The two distinct Hopf link textures can be distinguished by the structure of their potential PT surfaces. Using the right-hand rule, one can define an orientation for each link, and compute linking numbers. The $\textrm{Lk}=+1$ texture is also equivalent to the $\textrm{Lk}=-1$ texture with an additional tether connecting the two components. (d) The Pontryagin-Thom surface. The surface is constructed as the set of all points where the molecular orientation is perpendicular to a chosen direction ${\bf d}$; disclinations become boundaries of this surface.}
\label{fig:2}
\end{figure}

This classification on its own, while complete, does not give a physical interpretation of these states, nor does it prescribe a method to identify knots produced experimentally or in simulation. This is provided by the Pontryagin-Thom (PT) construction, illustrated in Fig.~\ref{fig:2}, which allows the different states to be distinguished by a combination of `Skyrmion tubes' and relative disclination orientation. This construction has been implemented experimentally~\cite{chen13} and its employment should enable the identification of knotted defects in the laboratory. The PT construction gives a rigorous and succinct way of viewing the global topological data encoded in a director field~\cite{BryanThesis,dieck08}. One draws the surface consisting of all points where the director is perpendicular to a chosen orientation ${\bf d}$. For a given ${\bf d}$, any $\pi$ rotation of the director will be perpendicular to ${\bf d}$ at least once, and so for line defects this construction produces a surface whose boundaries are disclination lines, shown in Fig.~\ref{fig:2}(d). An additional degree of freedom corresponding to the director orientation in the plane perpendicular to ${\bf d}$ is used to colour the surface, as in Fig.~\ref{fig:2}(b). While the PT surface need not be orientable, the direction in which the director rotates as one lifts off the surface defines a local orientation on the PT surface, which can be reversed through a $\pi$ rotation in the surface itself~\cite{note2}.

In terms of this construction, the difference between states is captured by differences in the topology and colouring of this surface. For a knot or link the PT construction produces a surface, $F$, whose boundary is $L$, demonstrated in Figs.~\ref{fig:2}(a) \& (c). This $F$ corresponds to a `branch-cut' between to the two sheets that was performed during the transition to $\widehat{\Sigma}(L)$. In Fig.~\ref{fig:2} this surface is shown in blue, the constant colour indicating that the orientation of the director on the surface is uniform. Textures also exist where the surface $F$ does \emph{not} have a constant colour. In these cases one can perform a deformation of the texture, corresponding to a cobordism of the surface, contracting the region of colour winding and pulling the coloured part of the surface away from $F$, until it takes the form of separate pieces of surface that are either tubes entangled with the surface as shown in Fig.~\ref{fig:2}(a), or tethers joining separate link components. These tubes or tethers have a $2\pi$ winding of the director around their meridian and can be thought of as `Skyrmion tubes', a baby Skyrmion, with a profile as in Fig.~\ref{fig:2}(b) extruded along a cycle. The cycle along which this tether or tube runs is precisely a cycle in $H_1(\widehat{\Sigma}(L),\partial \widehat{\Sigma}(L))$, giving a vivid visual correspondence between the classfication and physical realisation of these textures. 

An additional subtlety arises for links; $H_1(\Sigma(L))$ need not have a single generator for a multi-component link, so one needs to associate a class of tethers or tubes to each generator (for knots $H_1(\Sigma(L))=\mathbb{Z}_d$, where $d=\textrm{det}(L)$, and so knots have a single generator). Furthermore, links may support multiple topologically distinct planar textures, \textit{i.e.} the director lies everywhere in a single plane (the $(x$-$z)$ plane, say), which therefore do not possess any Skyrmion tubes. Fig.~\ref{fig:2}(c) furnishes an example of this, demonstrating representatives of the two distinct textures that may be associated to the Hopf link. These examples have interpretations as disclinations with linking number $\textrm{Lk}=\pm 1$. If the texture is planar then the local orientation of the PT surface may be extended globally and thus induces an orientation on the boundary components, \textit{i.e.} the disclination lines. The linking number of these oriented disclination lines can then be computed and used to distinguish between the states. 

\begin{figure}[t]
\begin{center}
\includegraphics{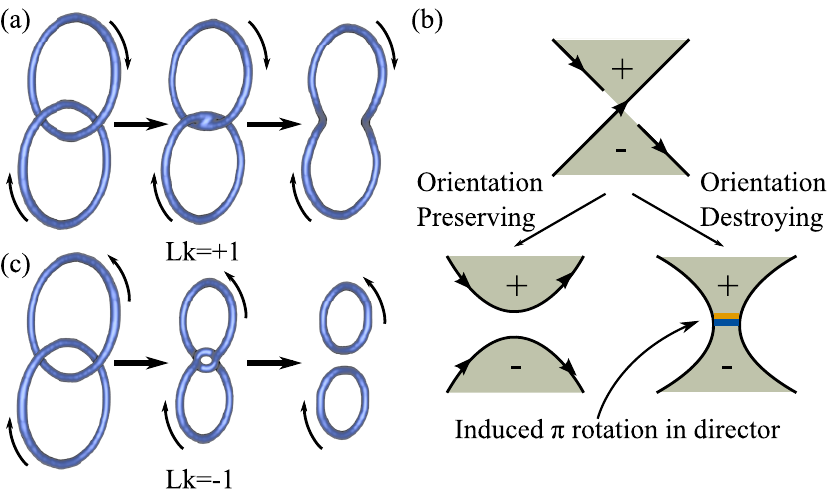}
\end{center}
\caption{The topology of defect crossing processes. (a) \& (c) Simulation results showing example relaxations of $\textrm{Lk}=+1$ and $\textrm{Lk}=-1$ Hopf link textures. The orientation induced by the PT surface, shown by black arrows, is preserved by the strand crossings. (b) Schematic of a possible defect strand cobordism. The Pontryagin-Thom surface is shown shaded. The orientation destroying resolution necessarily induces a $\pi$ rotation of the director in the surface, indicated by the coloured bar; this distortion is energetically unfavoured in a nematic and so in such systems the resolution will preserve the orientation induced by the PT surface.}
\label{fig:6}
\end{figure}

This topological distinction manifests itself in the physical behaviour of defect lines. A free-standing defect in a nematic will coarsen over time, undergoing strand crossings in order to reduce its free energy~\cite{ishikawa98,chuang91}. So while a pure nematic will not support stable knotted defects, we find that the behaviour and relaxation of unstable defects depends strongly on the topological class of the knot or link. We demonstrate this using the Hopf link, though the phenomenon is general in nature, with potential relevance to all defect crossing processes. Fig.~\ref{fig:6} shows simulated relaxation dynamics of two Hopf link defects, with linking numbers $\pm 1$.  The resolution of strand crossings in such relaxation processes is observed to \emph{always} preserve the effective orientation imparted to the defects by the Pontryagin-Thom surface (shown in Fig.~\ref{fig:2}(c)). This effect can be understood as a combination of energetics and topology. The simplification of defect lines can be pictured in terms of elementary cobordism moves, an example of which is shown in Fig.~\ref{fig:6}(b). These cobordism moves act not just on the defects themselves but on the defects \emph{and} the PT surface simultaneously, a reflection of the fact that the entire nematic texture is knotted, not just the disclinations. Locally the sides of the PT surface are distinguished, and for a cobordism to connect the two opposite sides (and thus facilitate a strand crossing) requires a local $\pi$ rotation in the director on the surface itself. This $\pi$ rotation is topologically required only if the cobordism does not preserve orientation. Since in a nematic one takes the energy to be a simple quadratic distortion energy, such distortions as produced by the orientation destroying cobordism are suppressed.

This behaviour has the potential to allow one to distinguish the topological types of links and knots purely by observation. For example, the two distinct Hopf links may be distinguished by the manner in which they annihilate, as in Fig.~\ref{fig:6}. More generally, this gives an initial characterisation of the topological dynamics of defects in nematic systems. Indeed, the outcome of generic defect processes, easily observed experimentally, can be predicted by this rule.

\begin{figure}[t]
\begin{center}
\includegraphics{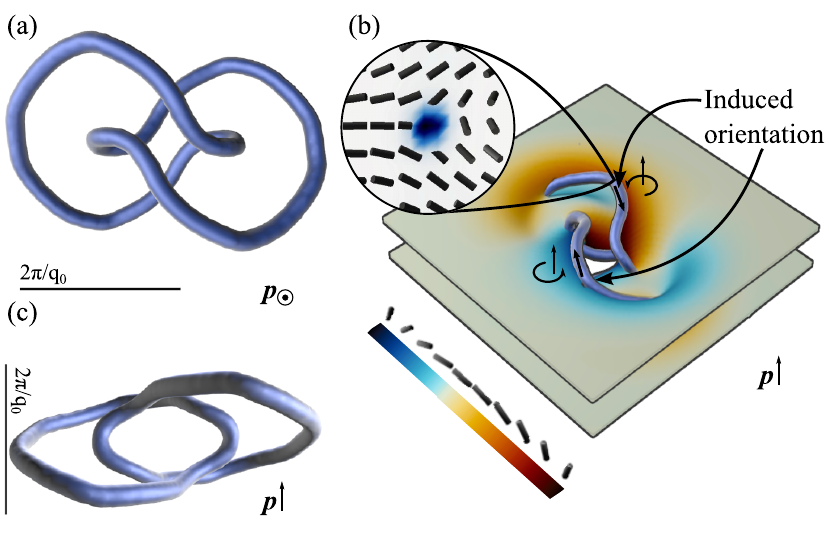}
\end{center}
\caption{Simulation results showing metastable $\textrm{Lk}=+1$ Hopf link in a cholesteric, found by numerical relxation of the Landau-de Gennes free energy. (a) \& (c) Vertical and horizontal views of the knotted configuration; the pitch direction is denoted by ${\bf p}$. Note that the size of the defects is comparable to $2\pi / q_0$. (b) PT surface plotted from simulation results. The two sheets correspond to a bulk cholesteric in which the link is embedded. The topological class of the link can be found by looking at the induced orientation on the link components by the PT surface and then computing the linking number. The legend at the bottom shows the relationship between surface colouring and director orientation. The inset shows the twist profile of the disclinations.}
\label{fig:4}
\end{figure}

What of the influence of topology on stable knotted configurations? The natural setting for stable knots is either in colloidal systems~\cite{tkalec11,martinez14,machon13,cavallaro13} or in cholesterics, where one can exploit the length scale set by the cholesteric pitch, as demonstrated in simulations of cholesteric droplets~\cite{sec14}. Here we show numerically the existence of a stable Hopf link in a bulk cholesteric, shown in Fig.~\ref{fig:4}. Using the PT construction we are able to extract the topological class of the configuration, demonstrating a possible experimental method of identification. Figs.~\ref{fig:4}(a) and (c) show views of a stable $\textrm{Lk}=+1$ Hopf link, while Fig.~\ref{fig:4}(b) shows the PT surface computed from the simulation results. The defect lines are of order one pitch length in size, and display a local twist profile in the director. The PT surface is orientable, and carries no colour winding, Skyrmion tethers or tubes. We may therefore use the induced orientation on the boundary components to compute the linking number of the defect lines and identify the state. While Fig.~\ref{fig:4} shows only a Hopf link, we have also been able to stabilise a large variety of knots and links, limited only by our persistence. The simulation shown has $q_0<0$ and it is interesting to note that the $\textrm{Lk}=-1$ Hopf link was only found to be stable with $q_0>0$, with a similar phenomenon shown for the left- and right-handed trefoil knots, suggesting a link between the chirality of the system and of the knot.

To make controlled simulations of these states, in both the nematic and cholesteric cases, we use Milnor fibrations~\cite{milnor68} to construct an anstaz containing a knotted director field. We take the complex polynomial $f(z_1,z_2)= z_1^p+(-iz_2)^q$ restricted to $S^3 \subset \mathbb{C}^2$, \textit{i.e.} $|z_1|^2+|z_2|^2=1$. Constructing the director ansatz 
\begin{equation}
n = \bigl(\cos(q_0z+ \phi/2), \sin(q_0z+\phi/2),0\bigr) ,
\label{eq:3}
\end{equation} 
where $\phi(\vec{x})=\textrm{Arg}(\pi_\ast f)$, the argument of the stereographic projection of $f$ into $\mathbb{R}^3$, gives a director field containing a defect in the form of a $(p,q)$ torus knot embedded into a standard cholesteric texture (or nematic for $q_0=0$). While the polynomial given only generates torus knots, an extension to other polynomials~\cite{dennis10,remond-tiedrez14} may be used to generate an entire zoo of knots and links. Furthermore the texture of~\eqref{eq:3} is planar, and cannot correspond to a topological class containing Skyrmion tubes; these additional topological classes may be produced by a generalisation involving appropriate meromorphic functions~\cite{machon13-2}. The topologically distinct planar textures for links may be constructed using conjugated polynomials~\cite{machon13-2,seade05}. For example, the choices $(z_1+z_2)(z_1-z_2)$ and $(z_1+z_2)\overline{(z_2-z_1)}$ give planar representatives of the $\textrm{Lk}=+1$ and $\textrm{Lk}=-1$ textures for the Hopf link, respectively~\cite{note3}. To simulate the system we use the standard Landau-de Gennes theory. We embed the director into a $Q$ tensor as $Q= s(n \otimes n - \frac{1}{3} \mathbb{I})$ and take the relaxation dynamics $\partial_t Q = - \Gamma \frac{\delta F}{\delta Q}$, where $F$ is the standard Landau-de Gennes free energy~\cite{ravnik09}. The simulations were run on a $192^3$ grid with typical parameter values. 

This work was partially supported by the EPSRC. TM also partially supported by a University of Warwick Chancellor's Scholarship.


\begin{thebibliography}{99}
\bibitem{note2} Technically the requirement is for a consistent map from the normal bundle of the PT surface into the normal bundle of the circle in $\mathbb{RP}^2$ defined by $\bf{d}$.
\bibitem{note3} Note that as polynomials they have the same zero set and thus disclinations with identical geometry.
\end{thebibliography}
\end{document}